\documentclass[11pt]{article}
\usepackage{amsmath}
\usepackage{amsfonts}

\usepackage{amsmath,amssymb}
\usepackage{graphicx}
\usepackage{color}
\usepackage{url}

\newtheorem{theorem}{Theorem}
\newtheorem{lemma}{Lemma}

\newtheorem{proposition}{Proposition}
\newtheorem{remark}{Remark}

\newtheorem{example}{Example}

\newcommand{\RNum}[1]{\uppercase\expandafter{\romannumeral #1\relax}}

\newcommand{\Tr}{{{\rm Tr}}}

\newcommand{\ls}[1]
    {\dimen0=\fontdimen6\the\font\lineskip=#1\dimen0
     \advance\lineskip.5\fontdimen5\the\font
     \advance\lineskip-\dimen0
     \lineskiplimit=0.9\lineskip
     \baselineskip=\lineskip
     \advance\baselineskip\dimen0
     \normallineskip\lineskip\normallineskiplimit\lineskiplimit
     \normalbaselineskip\baselineskip
     \ignorespaces}

\begin{document}

\bibliographystyle{abbrv}

\title{On the Correlation Distribution for a Ternary Niho Decimation}

\author{Yongbo Xia
\thanks{Y. Xia is with the Department of Mathematics and Statistics,
South-Central University for Nationalities, and is
  also with the Hubei Key Laboratory of Intelligent Wireless Communications,
  South-Central University for Nationalities, Wuhan 430074, China (e-mail:
xia@mail.scuec.edu.cn).},
 Nian Li
\thanks{N. Li  and T. Helleseth are with the  Department of Informatics, University of
Bergen, N-5020 Bergen, Norway (e-mail: Nian.Li@ii.uib.no;
Tor.Helleseth@ii.uib.no).}, Xiangyong Zeng
\thanks{X. Zeng is with the Faculty of Mathematics and Statistics, Hubei University, Wuhan 430062, China (e-mail:
xzeng@hubu.edu.cn).}, and Tor Helleseth \footnotemark[2]}
\date{}
\maketitle

\thispagestyle{plain} \setcounter{page}{1}

\begin{abstract}
In this paper, let $n=2m$ and $d=3^{m+1}-2$ with $m\geq2$ and $\gcd(d,3^n-1)=1$. By studying the weight distribution of the ternary Zetterberg code and counting the numbers of solutions of some equations over the finite field $\mathbb{F}_{3^n}$, the correlation distribution between a ternary $m$-sequence of period $3^n-1$ and its $d$-decimation sequence is completely determined. This is the first time that the correlation distribution for a non-binary Niho decimation has been determined since 1976.

\vspace{2mm}

\end{abstract}

\ls{1.5}
\section{Introduction}\label{sec1}
%
%
%
%

Let $p$ be a prime, $n$  a positive integer and $\{s(t)\}$  a $p$-ary $m$-sequence over the finite field $\mathbb{F}_p$ with period $p^n-1$.
A $d$-decimation sequence of $\{s(t)\}$ is given by $\{s(dt)\}$ and the integer $d$ is called a decimation.
The cross-correlation function $C_d(\tau)$ between $\{s(t)\}$ and its $d$-decimation sequence $\{s(dt)\}$ is defined by
\begin{equation}\label{dfC_d}
C_{d}(\tau)=\sum\limits_{t=0}^{p^n-2}\omega_p^{s\left(t+\tau\right)-s\left(dt\right)},
\end{equation}
where $\tau=0,\,1,\cdots,\,p^n-2$ and $\omega_p=e^{2\pi\sqrt{-1}\over p}$ is a primitive complex $p$-th
root of unity. In the theory of sequence design, for a decimation $d$ leading to low cross-correlation, it is interesting to determine the values of $C_d(\tau)$ together with the number of occurrences of each value, which is known as the correlation distribution for the decimation $d$. This problem has received a
lot of attention since the 1960s, and many interesting theoretical results have been obtained \cite{Gold1968,Trach,Niho,Helleseth76,Helleseth78,Hell-Kumar,Golomb,Rosen2004,feng2013,feng-ge2014, xiaieee2014}.  For known results and some open problems in this direction, the reader is referred to \cite[Section 2.2]{Rosen2004} and a recent survey paper \cite{Helleseth2014}.

A decimation $d$ is called a Niho decimation over the finite field $\mathbb{F}_{p^n}$ provided $n=2m$ for some positive integer $m$ and $$d\equiv p^i\,\,({\rm mod}\,\,p^m-1)$$
for some $i<n$.  The Niho decimation was originally studied by Niho in his famous thesis \cite{Niho}, and it leads to at least four-valued cross-correlation \cite{Charpin2004, Helleseth 2007niho}. Some basic properties about Niho decimations can be found in \cite{Dobbertin06,Rosen2004}, and for more research problems involving them, the reader is referred to \cite{Dobbertin98,Charpin2004,Rosen2004,Helleseth2005,Dobbertin06,Rosen2006,Helleseth 2007niho,Niho-BENT-Dobbertin06}. In the binary case $p=2$, all the known Niho decimations for which the correlation distributions are completely determined can be found in
the recent paper  \cite{xiaieee2016}. When $p$ is odd, there are only two such Niho decimations below:
\begin{quote}

(\textrm{i}) $d=2p^m-1$, where $n=2m$ and $p^m\not\equiv2\,\,({\rm mod}\,\,3)$ \cite[Theorem 4.3]{Helleseth76};

(\textrm{ii}) $d=\frac{p^n-1}{3}+p^s$, where $n\equiv 2\,\,({\rm mod}\,\,4)$, $0\leq s< n$, $p\equiv 2\,\,({\rm mod}\,\,3)$ and $\frac{1}{3}\,p^{n-s}\left(p^n-1\right)\not\equiv 2\pmod{3}$ \cite[Theorem 4.11]{Helleseth76}.
\end{quote}
 Case (\textrm{i}) leads to four-valued cross-correlation, while Case (\textrm{ii}) leads to six-valued cross-correlation. Note that both (\textrm{i}) and (\textrm{ii}) were found  by Helleseth in 1976, and since then no further results have been found.

Under the condition that $\gcd(5,p^m+1)=1$, the positive integer  \begin{equation}\label{deci studied}d=3p^m-2\end{equation} is coprime to $p^n-1$ and is a typical Niho decimation over the finite field $\mathbb{F}_{p^n}$, where $n=2m$. This decimation has been studied for a long time, and by the well-known Niho's theorem \cite{Niho,Rosen2004,Niho-BENT-Dobbertin06,Rosen2006}, it leads to at most six-valued cross-correlation. However, its  correlation distribution is not completely determined yet. In an unpublished manuscript dated 1999 \cite{Dobbertion unpublished}, Dobbertin, Helleseth and Martinsen proved that for $p=3$ the Niho decimation in (\ref{deci studied}) leads to five-valued cross-correlation, but they did not determine the correlation distribution.  Later, in 2006,  Dobbertin et al. published a result about the correlation distribution for the Niho decimation (\ref{deci studied}) in the binary case \cite{Dobbertin06}. Concretely, when $p=2$ and $m$ is odd, they expressed the correlation distribution for the Niho decimation (\ref{deci studied}) in terms of a class of exponential sums. Those exponential sums are involved in the Dickson polynomials and the Kloosterman sums, and  are generally difficult to be evaluated in closed forms.

Recently,  in \cite{xiaieee2016}, for the binary case $p=2$, the problem of determining the correlation distribution for the Niho decimation (\ref{deci studied}) was reduced to a combinatorial problem related to the unit circle of $\mathbb{F}_{2^n}$. Further, inspired by the idea in \cite{chenyuan2016}, the authors of \cite{xiaieee2016} established a connection between the combinatorial problem and the binary Zetterberg code, and then they determined the correlation distribution based on the weight distribution of the binary Zetterberg code in \cite{Zetterbergcodes91, Zetterberg-code07}. In the present paper, by using similar techniques of \cite{xiaieee2016}, we determine the correlation distribution for the Niho decimation (\ref{deci studied}) in the ternary case. Compared with \cite{xiaieee2016}, the procedure for establishing the connection here, however,  is much more complicated since the connection in the ternary case is not so direct and obvious as that in the binary case. In addition, there is
no weight formula available for the ternary Zetterberg code and thus we need to establish some of these formulas. This paper supplies a proof of a conjecture proposed by Dobbertin et al. in \cite{Dobbertion unpublished}. It is the first time that the correlation distribution for a non-binary Niho decimation has been determined since 1976.

The remainder of this paper is organized as follows. Section \ref{main result} states our main result on the correlation distribution for the  Niho decimation (\ref{deci studied}) in the ternary case. Section \ref{pre} introduces some notation and preliminaries.
Section \ref{proof for main} is devoted to proving our main result, and the concluding remarks are given in Section \ref{conclusion}.

\section{Main result on the correlation distribution}\label{main result}
Throughout the remainder of this paper, we always assume that $p=3$, \begin{equation}\label{def d}d=3^{m+1}-2,\end{equation}
and $n=2m$ with $m\geq 2$.  Let $\mathbb{F}_{3^n}^*$ denote the multiplicative group of the finite field $\mathbb{F}_{3^n}$. Define
\begin{equation}\label{def of Sab}
S(a,b)=\sum\limits_{x\in \mathbb{F}_{3^n}^*}\omega_3^{\Tr^n_1(ax+bx^{d})},
\end{equation}
where $(a,b)\in \mathbb{F}_{3^n}\times\mathbb{F}_{3^n}$, $\omega_3=e^{2\pi\sqrt{-1}\over 3}$ is a primitive complex $3$-th
root of unity, and $\Tr^n_1(\cdot)$ is the trace function from $\mathbb{F}_{3^n}$ to $\mathbb{F}_{3}$ \cite{LN}. The main results of this paper are given in Theorems \ref{exp-distribution theorem} and \ref{cor-distribution theorem} below.

\begin{theorem}\label{exp-distribution theorem} When $(a,b)$ runs through $\mathbb{F}_{3^n}\times\mathbb{F}_{3^n}$, the value distribution of $S(a,b)$ is given by
\begin{equation*}\label{valuedis-ternary}\left\{\begin{array}{rll}
3^n-1,&\,\, 1\,&\,{\rm time},\\
 -3^m-1,&\,\, \frac{\left(3^{2m} - 1\right)\left(11\cdot\,3^{2m} -16\cdot\,3^m -(-1)^m 3^m+6\right)}{30}\,&\,{\rm times},\\
  -1,&\,\frac{\left(3^{2m} - 1\right)\left(3\cdot\,3^{2m} + (-1)^m 3^m + 2\cdot\,3^m + 2\right)}{8}\,&\,{\rm times},\\
  3^m-1,&\,\,\frac{\left(3^{2m} - 1\right)\left(3^{2m} - (-1)^m 3^m + 6\right)}{6}\,&\,{\rm times},\\
  2\cdot3^m-1,&\,\, \frac{\left(3^{2m} - 1\right)\left(3^{2m} + (-1)^m 3^m + 4\cdot\,3^m - 6\right)}{12}\,&\,{\rm times},\\
      4\cdot3^m-1,&\,\,\frac{\left(3^{2m} - 1\right)\left(3^{2m} - 6\cdot\,3^m - (- 1)^m 3^m + 6\right)}{120} \,&\,{\rm times}.\end{array}\right.
\end{equation*}
\end{theorem}

In order to make sure $\gcd(3^{m+1}-2,3^n-1)=1$, it requires that $\gcd(5,3^m+1)=1$, which is equivalent to $m\not\equiv2\,\,({\rm mod}\,\,4)$. As a consequence of Theorem \ref{exp-distribution theorem}, the correlation distribution for the ternary Niho decimation (\ref{def d}) can be derived immediately.

\begin{theorem}\label{cor-distribution theorem}  Let $n=2m$ and $d=3^{m+1}-2$ with $m\not\equiv2\,\,({\rm mod}\,\,4)$. Then, the distribution of the cross-correlation function $C_d(\tau)$ defined in (\ref{dfC_d}) is given by
\begin{equation*}\label{valuedis-ternary}\left\{\begin{array}{rll}
 -3^m-1,&\,\, \frac{11\cdot\,3^{2m} -16\cdot\,3^m -(-1)^m 3^m+6}{30}\,&\,{\rm times},\\
  -1,&\,\frac{3\cdot\,3^{2m} + (-1)^m 3^m + 2\cdot\,3^m -14}{8}\,&\,{\rm times},\\
  3^m-1,&\,\,\frac{3^{2m} - (-1)^m 3^m + 6}{6}\,&\,{\rm times},\\
  2\cdot3^m-1,&\,\, \frac{3^{2m} + (-1)^m 3^m + 4\cdot\,3^m - 6}{12}\,&\,{\rm times},\\
      4\cdot3^m-1,&\,\,\frac{3^{2m} - 6\cdot\,3^m - (- 1)^m 3^m + 6}{120} \,&\,{\rm times}.\end{array}\right.
\end{equation*}
\end{theorem}

The proofs of Theorems \ref{exp-distribution theorem}  and \ref{cor-distribution theorem} are presented in Section \ref{proof for main}. In what follows, we provide two examples to verify our results.
\begin{example} Let $m=3$, then $n=2m=6$ and $d=3^{m+1}-2=79$. Let $\alpha$ be a primitive element of the finite field $\mathbb{F}_{3^6}$ and
$\{s(t)\}$ be a ternary $m$-sequence given by $s(t)={\Tr}_1^6(\alpha^t)$. By Magma, the value distribution of $S(a,b)$ is given as follows
\begin{displaymath}\left\{\begin{array}{rcl}
728,&\,\,1\,&{\rm time},\\
 -28,&\,\,184912\,&{\rm times},\\
  -1,&\,\,201656\,&{\rm times},\\
 26,&\,\, 92456\,&{\rm times},\\
 53,&\,\,48776\,&{\rm times},\\
  107,&\,\,3640\,&{\rm times}. \end{array}\right.
\end{displaymath}
The distribution of the cross correlation function $C_d(\tau)$ between $\{{\Tr}_1^6\left(\alpha^t\right)\}$ and its decimation sequence $\{{\Tr}_1^6\left(\alpha^{dt}\right)\}$ is given by
\begin{displaymath}\left\{\begin{array}{rcl}
 -28,&\,\,254\,&{\rm times},\\
  -1,&\,\,275\,&{\rm times},\\
 26,&\,\, 127\,&{\rm times},\\
 53,&\,\,67\,&{\rm times},\\
  107,&\,\,5\,&{\rm times}. \end{array}\right.
\end{displaymath}

\end{example}

\begin{example} Let $m=4$, $n=2m=8$, $d=3^{m+1}-2=241$, and $\{s(t)\}$ be the ternary $m$-sequence given by $s(t)={\Tr}_1^8(\alpha^t)$, where $\alpha$ is a primitive element of the finite field $\mathbb{F}_{3^8}$.  Then, by Magma, the value distribution of $S(a,b)$ is
\begin{displaymath}\left\{\begin{array}{rcl}
 6560,&\,\,1\,&{\rm time},\\
 -82,&\,\, 15481600\,&{\rm times},\\
  -1,&\,\,16340960\,&{\rm times},\\
 80,&\,\, 7091360\,&{\rm times},\\
 161,&\,\,3804800\,&{\rm times},\\
  323,&\,\,328000\,&{\rm times},\end{array}\right.
\end{displaymath}
and the distribution of the cross-correlation function $C_d(\tau)$ between $\{s(t)\}$ and $\{s(dt)\}$ is given by
\begin{displaymath}\left\{\begin{array}{rcl}
 -82,&\,\,2360\,&{\rm times},\\
  -1,&\,\,2489\,&{\rm times},\\
 80,&\,\, 1081\,&{\rm times},\\
 161,&\,\,580\,&{\rm times},\\
  323,&\,\,50\,&{\rm times}. \end{array}\right.
\end{displaymath}
\end{example}

The above numerical results are coincided with the results in Theorems \ref{exp-distribution theorem}  and \ref{cor-distribution theorem}.

\section{Preliminaries}\label{pre}
In this section, we introduce some notation and preliminaries. For convenience, sometimes we denote $3^m$ by $q$.
Let $\alpha$ be a fixed primitive element of the finite field $\mathbb{F}_{3^n}$, then $\gamma=\alpha^{q+1}$ is a primitive element of $\mathbb{F}_{q}$. The unit circle of $\mathbb{F}_{3^n}$ is defined by
 \begin{equation}\label{Du}U=\left\{\,x \in \mathbb{F}_{3^n} \,\mid\, x \bar{x}=1\,\right\}, \end{equation}where $\bar{x}=x^{q}$. Note that $U$ is a cyclic subgroup of order $q+1$ in the multiplicative group $\mathbb{F}_{3^n}^*$. Actually,  $U=\left\{\, \alpha^{i(q-1)} \,\mid\, i=0,1,\cdots, q\,\right\}$ and $\alpha^{q-1}$ is a generator of $U$. Let\begin{equation}\label{Define omega}\Omega=\{\alpha^i:\,\,i=0,1,\cdots,q\},\end{equation} then every $x\in \mathbb{F}_{3^n}^*$ has a unique representation
\begin{equation}\label{polar rep}x=\beta y\end{equation} with $(\beta,y)\in \Omega \times \mathbb{F}_{q}^*$.

\subsection{Ternary Melas codes and ternary Zetterberg codes}
Let $m\geq2$ and $\gamma=\alpha^{q+1}$ be the primitive element of $\mathbb{F}_{q}$. Denote the minimal polynomial of $\gamma^i$ over $\mathbb{F}_{3}$ by $m_i(x)$, where $i\in \{-1,1\}$. The ternary Melas code $M(q)$ is the cyclic code over $\mathbb{F}_{3}$ of length $q-1$ generated by $m_1(x)m_{-1}(x)$, and its dual  code $M(q)^{\perp}$ is given by \cite{ternarymelascode,ternarymelaszetter}
\begin{equation}
\left\{\left(\Tr_1^m(ax+b/x)\right)_{x\in \mathbb{F}_{q}^*}:\,a,b\in \mathbb{F}_{q}\right\}.
\end{equation}
For each $i\in \{0,1,\cdots,q\}$, let $A_i$ denote the number of codewords of weight $i$ in $M(q)$. In \cite[Theorem 2.3]{ternarymelaszetter}, a formula for $A_i$ was derived, but the formula involves the traces of Hecke operators on certain spaces of cusp forms. Thus, for given $i$ and $m$, it is difficult to compute the value of $A_i$ explicitly. Later, in \cite{ternarymelascode},  the formula for $A_i$ was further illustrated and a table of formulas for $A_i$ with small $i$ was computed.  In the following lemma, we give the first five formulas in the table, which are useful in this paper.

\begin{lemma}\cite[Table 6.1]{ternarymelascode}\label{weightofMelas} With the notation introduced above, we have
\begin{equation}\label{Nvalue}\left \{
\begin{array}{cll}A_1&=&0,\\
A_2&=&q-1,\\
A_3&=&0,\\
A_4&=&\frac{(q-1)(q-3)}{2},\\
A_5&=&\frac{4(q-1)\left(q^2+\left((-1)^m-14\right)q+36\right)}{15}.\\
\end{array}\right.
\end{equation}
\end{lemma}

Let $\delta$ be a generator of $U$. For $m\geq 2$, the ternary Zetterberg code $Z(q)$ is a cyclic code over $\mathbb{F}_{3}$ of length $q+1$ defined by
 \begin{equation}\label{terZetterberg code}
 Z(q)=\left\{\left(c_0,c_1,\cdots,c_{q}\right)\in \mathbb{F}_{3}^{q+1}\mid \sum\limits_{i=0}^{q}c_i\delta^{i}=0\right\}.
 \end{equation}
The dual code $ Z(q)^{\perp}$ of $Z(q)$ has a very simple trace description \cite{ternarymelaszetter}:
 \begin{equation*}\label{dual of Zetter}
  Z(q)^{\perp}=\left\{\left(\Tr_1^n(a\delta^0),\Tr_1^n(a\delta),\cdots, \Tr_1^n(a\delta^{q})\right):\,a\in \mathbb{F}_{3^n}\right\}.
 \end{equation*} Let $B_i$ denote the number of codewords with weight $i$ in $Z(q)$, $0\leq i\leq q+1$. Let $$A_{M}(z)=\sum\limits_{i=0}^{q-1}A_iz^i\,\,{\rm and}\,\,A_{Z}(z)=\sum\limits_{i=0}^{q+1}B_iz^i$$ be the weight enumerators of $M(q)$ and $Z(q)$, respectively. In \cite{ternarymelaszetter}, by establishing a correspondence between the codewords of $M(q)^{\perp}$ and those of $ Z(q)^{\perp}$, the following result is deduced.

 \begin{lemma}\cite[pp. 268]{ternarymelaszetter}\label{AM-AZ} With the notation above, let $q'=\frac{q}{3}$, then we have \begin{equation}\label{enumerator-relation}\begin{array}{cll} &&\frac{q^2}{q+1}(1+2z)^{q'-1}A_{Z}(z)\\
 &=&\frac{q^2}{q-1}(1-z)^{q'+1}(1+z)^{q-1}A_M\left(\frac{-z}{z+1}\right)\\
 &&-2(1+2z)^{2q'}(1-z)^{2q'}-\frac{1}{q-1}(1-z)^{4q'}+\frac{1}{q+1}(1+2z)^{4q'}.\end{array}\end{equation}

  \end{lemma}

Combining Lemmas \ref{weightofMelas} and \ref{AM-AZ}, we can derive some weight formulas for $Z(q)$ as follows, which will play an important role in proving our main result in the sequel.
\begin{lemma}\label{weightofzetterberg}Let $B_i$ denote the number of codewords with weight $i$ in the ternary Zetterberg code $Z(q)$, then
$$
\left\{\begin{array}{cll}B_0&=& 1,\\
B_1&=& 0,\\
B_2&=& 3^m + 1,\\
B_3&=& 0,\\
B_4&=& \frac{\left(3^{2m}-1\right)}{2},\\
B_5&=& \frac{4\left(3^m + 1\right)\left(3^{2m} - 6\cdot3^m - {(- 1)}^m3^m + 6\right)}{15}. \end{array}\right.
$$
\end{lemma}
{\em Proof:} For each $i$ satisfying $0\leq i\leq q+1$, by comparing the coefficients of $z^i$
 on both sides of (\ref{enumerator-relation}), we have
 \begin{equation*}\label{equality for coef}\begin{array}{cll} &&
 \frac{q^2}{q+1}\sum\limits_{j=0}^i\left(\begin{array}{c}
q'-1 \\
i-j
\end{array}
\right)2^{i-j}B_j\\
&=&\frac{q^2}{q-1}\sum\limits_{j=0}^i\left[\left(\begin{array}{c}
q'+1 \\
j
\end{array}
\right)(-1)^j\sum\limits_{k=0}^{i-j}A_k(-1)^k\left(\begin{array}{c}
q-1-k \\
k
\end{array}
\right)\right]\\
&&-2\sum\limits_{j=0}^i\left(\begin{array}{c}
2q' \\
j
\end{array}
\right)2^j\left(\begin{array}{c}
2q' \\
i-j
\end{array}
\right)(-1)^{i-j}\\
&&-\frac{1}{q-1}\left(\begin{array}{c}
4q' \\
i
\end{array}
\right)(-1)^{i}+\frac{1}{q+1}\left(\begin{array}{c}
4q' \\
i
\end{array}
\right)2^{i},\end{array}
 \end{equation*}
where $\left(\begin{array}{c}
u \\
v
\end{array}
\right)$ denotes the number of $v$-combinations of an $u$-set. In the above equation, let $i$ take $0,1,\cdots,5$, respectively, then a system of linear equations in variables $B_0,B_1,\cdots, B_5$ is established. Solving this system gives the desired result.
\hfill$\square$

\subsection{Some combinatorial problems related to the unit circle}
Let $U$ be the unit circle of $\mathbb{F}_{3^n}$ defined in (\ref{Du}) and $\delta$ be a generator of $U$. Define a set
\begin{equation}\label{k-tuple of U}\mathcal{T}_k=\left\{(t_1,t_2,\cdots,t_k)\mid \delta^{t_1}+\delta^{t_2}+\cdots+\delta^{t_k}=0\,\,{\rm with}\,\,0\leq t_1<t_2<\cdots<t_k\leq q\right\},\end{equation} where $k$ is a positive integer and $k\geq 2$. Denote the cardinality of $\mathcal{T}_k$ by $\mid\mathcal{T}_k\mid$. For large $k$,  determining $\mid\mathcal{T}_k\mid$ is a very complicated problem. Due to Lemma \ref{weightofzetterberg}, $\mid\mathcal{T}_k\mid$ can be computed for $k=3,4,5$.

\begin{lemma}\label{threesum}With the notation above, we have $\mid\mathcal{T}_3\mid=0$.\end{lemma}
{\em Proof:} We only need to show that $\mathcal{T}_3$ is an empty set. Suppose, on the contrary, that there exists an element $(t_1,t_2,t_3)$ in $\mathcal{T}_3$. Then, we can define a vector $\mathbf{c}=(c_0,c_1,\cdots,c_{3^m})$ with
\begin{equation*}c_{t_1}=c_{t_2}=c_{t_3}\in \mathbb{F}_{3}^*\,\,{\rm and}\,\,c_i=0\,\,{\rm for}\,\,i\notin \{t_1,t_2,t_3\}.\end{equation*}
From the definition of $Z(q)$ in (\ref{terZetterberg code}) and the fact that $\delta^{t_1}+\delta^{t_2}+\delta^{t_3}=0$, it follows that
the vector $\mathbf{c}$ defined above is a codeword of weight three in $Z(q)$, a contradiction to Lemma \ref{weightofzetterberg} which states that there is no codeword with weight three in $Z(q)$. Thus, $\mathcal{T}_3$ is empty.
\hfill$\square$

\begin{lemma}\label{four U}With the notation introduced above, $\mathcal{T}_4$ is exactly given by \begin{equation}\label{condition for four u}\left\{(t_1,t_2,t_3,t_4)\mid 0\leq t_1<t_2<\frac{3^m+1}{2}, \,\,t_3=t_1+\frac{3^m+1}{2}\,\,{\rm and}\,\,t_4=t_2+\frac{3^m+1}{2}\right\}\end{equation}
and thus $\mid\mathcal{T}_4\mid=\frac{3^{2m}-1}{8}$.\end{lemma}
{\em Proof:}
Note that $\delta^{\frac{3^m+1}{2}}=-1$. If $(t_1,t_2,t_3,t_4)$ is an element of (\ref{condition for four u}), it is easily seen that $\delta^{t_1}+\delta^{t_2}+\delta^{t_3}+\delta^{t_4}=0$. Hence, every element of (\ref{condition for four u}) is also an element of $\mathcal{T}_4$. In the sequel, it suffices to show that except for the elements of (\ref{condition for four u}), there is no other $4$-tuple $(t_1,t_2,t_3,t_4)$ belonging to $\mathcal{T}_4$.

For each given element $(t_1,t_2,t_3,t_4)$ of (\ref{condition for four u}) ( also an element in $\mathcal{T}_4$), let $\mathbf{c}=(c_0,c_1,\cdots,c_{q})$ be a vector satisfying \begin{equation}\label{codeword-weight4}c_{t_1}=c_{t_3}\in \mathbb{F}_{3}^*,\,\, c_{t_2}=c_{t_4}\in \mathbb{F}_{3}^*\,\,{\rm and}\,\,c_i=0\,\,{\rm for}\,\,i\notin\{t_1,t_2,t_3,t_4\}.\end{equation}
Then $\mathbf{c}$ is a codeword with weight four in $Z(q)$. Note that for each given element $(t_1,t_2,t_3,t_4)$ of (\ref{condition for four u}), there are four different vectors $\mathbf{c}$ satisfying (\ref{codeword-weight4}). Moreover, for different elements $(t_1,t_2,t_3,t_4)$ of (\ref{condition for four u}), the corresponding codewords defined by (\ref{codeword-weight4}) are also different.
Thus, corresponding to the elements in (\ref{condition for four u}), there are $$\frac{3^{2m}-1}{8}\times 4=\frac{3^{2m}-1}{2}$$
different codewords with weight four in $Z(q)$ since the total number of elements $(t_1,t_2,t_3,t_4)$ in (\ref{condition for four u}) is $\frac{(1+\frac{3^m-1}{2})\frac{3^m-1}{2}}{2}=\frac{3^{2m}-1}{8}.$

Suppose that there exists an element $(t_1,t_2,t_3,t_4)$ in  $\mathcal{T}_4$ but not in (\ref{condition for four u}). Then, such an element $(t_1,t_2,t_3,t_4)$ gives at least two codewords $\mathbf{c}=(c_0,c_1,\cdots,c_{3^m})$ with weight four in $Z(q)$, which are given by
\begin{displaymath}c_{t_1}=c_{t_2}=c_{t_3}=c_{t_4}\in \mathbb{F}_{3}^*\,\,{\rm and}\,\,c_i=0\,\,{\rm for}\,\,i\notin \{t_1,t_2,t_3,t_4\}.\end{displaymath} Thus, except for the elements $(t_1,t_2,t_3,t_4)$ of (\ref{condition for four u}), if there exists other $4$-tuples $(t_1,t_2,t_3,t_4)$ in $\mathcal{T}_4$, the total number of codewords having weight four in $Z(q)$ will be greater than $\frac{3^{2m}-1}{2}$. So we arrive at a contradiction since  by Lemma \ref{weightofzetterberg}, the total number of codewords with weight four in $Z(q)$ is exactly $\frac{3^{2m}-1}{2}$. Therefore, except for the elements of (\ref{condition for four u}), there is no other $4$-tuple $(t_1,t_2,t_3,t_4)$ in  $\mathcal{T}_4$.

The cardinality of $\mathcal{T}_4$ is equal to the number of elements in (\ref{condition for four u}), which is $\frac{3^{2m}-1}{8}$.
\hfill$\square$

\begin{lemma}\label{five-sum-num} With the notation of Lemma \ref{weightofzetterberg}, let $\mathcal{T}_5$ be defined by (\ref{k-tuple of U}), then $\mid\mathcal{T}_5\mid=B_5/2^5$.\end{lemma}
{\em Proof:} See Appendix A.
\hfill$\square$

\section{Proof for the main result}\label{proof for main}
In this section, we will give the proof for the main result in Section \ref{main result}. Let $S(a,b)$ be the exponential sum defined in
(\ref{def of Sab}). The possible values of $S(a,b)$ given in Lemma \ref{Niho theorem of Sab} below can be found by the techniques used in Lemma 2 of \cite{linian2013}, which originate from the proof of Niho's Theorem \cite{Niho}.
\begin{lemma}\label{Niho theorem of Sab}Let $S(a,b)$ be the exponential sum defined in (\ref{def of Sab}). Then, the value of $S(a,b)$ is given by
\begin{equation*}(N(a,b)-1)3^m-1,\end{equation*}
where $N(a,b)$ is the number of $z\in U$ such that
\begin{equation}\label{eq-in lem1}bz^5+\bar{a}z^3+az^2+\bar{b}=0.\end{equation}
\end{lemma}

Note that (\ref{eq-in lem1}) has at most five roots in $U$ since its degree is at most five. Thus, the possible values of $S(a,b)$ are $3^n-1$, $-3^m-1$, $-1$, $2\cdot 3^m-1$, $3\cdot 3^m-1$, and $4\cdot 3^m-1$. In particular, $S(a,b)=3^n-1$ if and only if $a=b=0$. Moreover, the following lemma can exclude a redundant value from the possible values of $S(a,b)$. A similar result was already presented in \cite[Theorem 3.7]{Rosen2004}. For the reader's convenience, we include a proof here.

\begin{lemma}\label{exclude value}Let $S(a,b)$ be the exponential sum defined in (\ref{def of Sab}). Then, $S(a,b)\neq 3\cdot 3^m-1 $ for any $(a,b) \in \mathbb{F}_{3^n}\times \mathbb{F}_{3^n}$. \end{lemma}
{\em Proof:} By Lemma \ref{Niho theorem of Sab}, it suffices to prove that (\ref{eq-in lem1}) cannot have four roots in $U$. Now assume that
 (\ref{eq-in lem1}) has four roots in $U$. Then, $\bar{b}\neq 0$, ${{\bar{b}}/b}\in U$, and the fifth root of (\ref{eq-in lem1}) is also in $U$. Thus, when (\ref{eq-in lem1}) has four roots in $U$,
 it must have three roots with multiplicity $1$ and one root with multiplicity $2$. Therefore, the derivative
  \begin{equation}\label{der}2bz^4+2az\end{equation}
  of $bz^5+\bar{a}z^3+az^2+\bar{b}$ has a common root with (\ref{eq-in lem1}). Then, $a\neq 0$ and the only nonzero root of (\ref{der}) satisfies
  \begin{equation}\label{der1}z^3=-a/b.\end{equation}
  Substituting (\ref{der1}) into (\ref{eq-in lem1}), we get
  \begin{equation*}\bar{a}/\bar{b}=b/a,\end{equation*}
  which further implies (\ref{eq-in lem1}) has the following factorization
  \begin{equation*}b(z^3+a/b)(z^2+\bar{b}/a)=0.\end{equation*} The above factorization shows (\ref{eq-in lem1}) has at most three roots in
   $\mathbb{F}_{3^n}$, a contradiction. Thus, (\ref{eq-in lem1}) cannot have four roots in $U$. The desired conclusion is obtained.
\hfill$\square$

By Lemmas \ref{Niho theorem of Sab} and \ref{exclude value}, the nontrivial values of $S(a,b)$ are $(i-1)3^m-1$, $i=0,1,2,3,5$.  When $(a,b)$ runs through $\mathbb{F}_{3^n}^2\setminus\{(0,0)\}$, let $\mu_i$ denote the number of occurrences of $(i-1)3^m-1$, where $i=0,1,2,3$ and $\mu_4$ denote the number of occurrences of $4\cdot 3^m-1$. Determining the value distribution of $S(a,b)$ is exactly determining the values of $\mu_i$, $i=0,1,2,3,4$. In order to determine $\mu_i$, sufficient independent equations in terms of $\mu_i$'s should be obtained, and an efficient way to get these equations is computing the power sums of $S(a,b)$.
The following lemma is very useful for computing the power sums of $S(a,b)$. Its proof is routine and is omitted here.

\begin{lemma}\label{powersum} Let $N_r$ denote the number of solutions of

\begin{equation}\label{system of eq2}\left \{
\begin{array}{cll}x_1+x_2+\cdots+x_r=0,\\
x_1^{d}+x_2^{d}+\cdots+x_r^{d}=0,\\
\end{array}\right.
\end{equation}
in $\left(\mathbb{F}_{3^n}^*\right)^r$, where $d$ is given in (\ref{def d}), then we have
\begin{equation}\label{eq in lem2}\sum\limits_{(a,b)\in \mathbb{F}_{3^n}^2 }S(a,b)^r=3^{2n}N_r.\end{equation}
\end{lemma}

Note that $S(a,b)$ has five nontrivial values. In order to get five independent equations in terms of $\mu_i$'s, $i=0,1,\cdots,4$, we need to determine $N_r$ for $r=1,2,3,4$. When $r\geq 3$, it is usually difficult to determine the value $N_r$. In \cite{Xiong2014}, the authors introduced an elegant method for counting the number of solutions of certain equation systems related to generalized Niho decimations. Their main idea is transforming the equation system similar to (\ref{system of eq2}) into a matrix equation based the polar coordinate representations of the elements  in the finite fields. Combining this idea and Lemma \ref{five-sum-num}, the values of $N_3$ and $N_4$ can be determined as follows.

\begin{proposition}\label{Npro}With the notation above, we have
\begin{equation*}\left \{
\begin{array}{cll}N_1&=&0,\\
N_2&=&3^n-1,\\
N_3&=&(3^n-1)(3^m-2),\\
N_4&=&\left(3^{n}-1\right)\left(5\cdot 3^{n} - 12\cdot 3^m - (-1)^m\cdot 3^m + 9\right).
\end{array}\right.
\end{equation*}
\end{proposition}
{\em Proof:} It is easy to see that $N_1=0$ and $N_2=3^n-1$. Thus, we only consider the cases $r=3$ and $4$.

\emph{Determining $N_3$.} Let $N_3'$ be the number of solutions of
\begin{equation}\label{eq system for N3 tran0}\left \{
\begin{array}{lll}x_1+x_2=-1,\\
x_1^{d}+x_2^{d}=-1.\\
\end{array}\right.
\end{equation}
Then, \begin{equation}\label{relation btw N3 and N3' }N_3=(3^n-1)N_3'.\end{equation}
Let $(x_1,x_2)\in \left(\mathbb{F}_{3^n}^*\right)^2$. By (\ref{polar rep}), each $x_j$ can be represented as
$x_j=\beta_j y_j$ with $y_j\in \mathbb{F}_{3^m}^*$ and $\beta_j\in \Omega$,
where $j=1,2$ and $\Omega$ is defined in (\ref{Define omega}).
Then, \begin{equation}\label{subs-0}
x_j^{d}=\beta_j^{d}y_j=\beta_j^{3(3^m-1)+1}y_j=u_j^{3}\beta_jy_j,
\end{equation}
where $u_j=\beta_j^{3^m-1}$.
By the substitution in (\ref{subs-0}), we can write (\ref{eq system for N3 tran0}) as a matrix equation
\begin{equation}\label{eq symtem matrix1}
\left(
  \begin{array}{cccc}
    1 & 1 & \\
    u_1^3 & u_2^3 \\
  \end{array}
\right)
\cdot
\left(
  \begin{array}{c}
    \beta_1 y_1  \\
    \beta_2 y_2  \\
      \end{array}
\right)
=
\left(
  \begin{array}{c}
   -1 \\
  -1  \\
  \end{array}
\right).
\end{equation}
Therefore, $N_3'$ is the number of $\left(\beta_1,\beta_2,y_1,y_2\right)\in \Omega^2\times \left(\mathbb{F}_{3^m}^*\right)^2$ satisfying (\ref{eq symtem matrix1}).
Let
 \begin{equation*}\label{Coffmatrix}
A=\left(
  \begin{array}{cccc}
    1 & 1\\
    u_1^{3} & u_2^{3}\\
  \end{array}
\right).
\end{equation*}
Then,
\begin{equation*}\label{detA}
\det(A)=u_2^3 - u_1^3=\left(u_2-u_1\right)^3.
\end{equation*}
Now we determine the solutions of (\ref{eq symtem matrix1}) according to the determinant of the matrix $A$.

\emph{Case 1:} $\det(A)=0$. Then, $u_1=u_2$, which implies $\beta_1=\beta_2$, and (\ref{eq symtem matrix1}) becomes
\begin{equation*}\label{ N3 tran1}\left \{
\begin{array}{lll}\beta_1=\beta_2,\\
\beta_1y_1+\beta_2y_2=-1,\\
u_1^3\left(\beta_1y_1+\beta_2y_2\right)=-1,\\
\end{array}\right.
\end{equation*}which is equivalent to
\begin{equation}\label{ N3 tran2}\left \{
\begin{array}{lll}\beta_1=\beta_2=1,\\
y_1+y_2=-1.\\
\end{array}\right.
\end{equation}The number of $(\beta_1,\beta_2,y_1,y_2)\in \Omega^2\times \left(\mathbb{F}_{3^m}^*\right)^2$ satisfying (\ref{ N3 tran2}) is $3^m-2$.

\emph{Case 2:} $\det(A)\neq 0$. Then, $u_1\neq u_2$. For each given $(\beta_1,\beta_2)\in \Omega^2$ such that $u_1\neq u_2$, from (\ref{eq symtem matrix1}), we can solve a unique solution $(y_1,y_2)$ as follows
\begin{equation}\left \{
\begin{array}{lll}\beta_1y_1=\frac{u_2^3-1}{u_1^3-u_2^3},\\
\beta_2y_2=-\frac{u_1^3-1}{u_1^3-u_2^3}.\\
\end{array}\right.
\end{equation}
To ensure $y_i\in \mathbb{F}_{3^m}^*$, $i=1,2$, we must have
\begin{equation*}\left \{
\begin{array}{lll}u_1=\left(\frac{u_2^3-1}{u_1^3-u_2^3}\right)^{3^m-1},\\
u_2=\left(-\frac{u_1^3-1}{u_1^3-u_2^3}\right)^{3^m-1},\\
\end{array}\right.
\end{equation*}
which implies $u_1=u_2$, a contradiction. Thus, when $\det(A)\neq 0$, (\ref{eq symtem matrix1}) has no solution.

Combining Cases 1 and 2, we have $N_3'=3^m-2$. From (\ref{relation btw N3 and N3' }), the desired result follows.

\emph{Determining $N_4$.} Similarly, let $N_4'$ be the number of solutions of
\begin{equation}\label{eq system for N4 tran0}\left \{
\begin{array}{lll}x_1+x_2+x_3=-1,\\
x_1^{d}+x_2^{d}+x_3^d=-1.\\
\end{array}\right.
\end{equation}
Then, \begin{equation}\label{relation btw N4 and N4' }N_4=(3^n-1)N_4'.\end{equation}
Note that (\ref{eq system for N4 tran0}) is equivalent to

\begin{equation}\label{eq system for N4 tran1}\left \{
\begin{array}{lll}x_1+x_2+x_3=-1,\\
x_1^{3^m}+x_2^{3^m}+x_3^{3^m}=-1,\\
x_1^{d}+x_2^{d}+x_3^d=-1.\\
\end{array}\right.
\end{equation}
Using the substitution in (\ref{subs-0}) and noting that $x_j^{3^m}=u_j\beta_jy_j$, where $u_j=\beta_j^{3^m-1}$ and $j=1,2,3$,  (\ref{eq system for N4 tran1}) can be written as
\begin{equation}\label{system matrix for r=4}
\left(
  \begin{array}{cccc}
   1 & 1 &  1\\
    u_1 & u_2 & u_3 \\
    u_1^{3} & u_2^{3} &  u_3^{3}\\
        \end{array}
\right)
\cdot
\left(
  \begin{array}{c}
    \beta_1 y_1  \\
    \beta_2 y_2  \\
   \beta_3 y_3  \\
    \end{array}
\right)
=
\left(
  \begin{array}{c}
    -1  \\
   -1  \\
   -1 \\
    \end{array}
\right).
\end{equation}
Then, $N_4'$ is the number of $\left(\beta_1,\beta_2,\beta_3,y_1,y_2,y_3\right)\in \Omega^3\times \left(\mathbb{F}_{3^m}^*\right)^3$ satisfying (\ref{system matrix for r=4}).
Let \begin{equation*}\label{ N4Coffmatrix}
B=\left(
  \begin{array}{cccc}
   1 & 1 &  1\\
    u_1 & u_2 & u_3 \\
    u_1^{3} & u_2^{3} &  u_3^{3}\\
        \end{array}
\right).
\end{equation*}
Then,
\begin{equation*}\label{detB}
\det(B)=-\left(u_1 - u_2\right)\left(u_1 - u_3\right)\left(u_2 - u_3\right)\left(u_1 + u_2 + u_3\right).
\end{equation*}

\emph{Case \textrm{I}:} $\det(B)=0$. By Lemma \ref{threesum}, $u_1+u_2+u_3=0$ if and only if $u_1=u_2=u_3$. Thus, we consider the following cases.

\emph{Subcase (a):} $u_1=u_2\neq u_3$. Then, $\beta_1=\beta_2\neq \beta_3$ and (\ref{system matrix for r=4}) becomes
\begin{equation}\label{eq system for N4 tran2}\left \{
\begin{array}{lll}\beta_1=\beta_2\neq\beta_3,\\
\beta_1( y_1+ y_2)+\beta_3 y_3=-1,\\
u_1^3\beta_1( y_1+ y_2)+u_3^3\beta_3 y_3=-1.\\
\end{array}\right.
\end{equation}
If $y_1+y_2=0$, then (\ref{eq system for N4 tran2}) is transformed into
\begin{equation}\label{eq system for N4 tran3}\left \{
\begin{array}{lll}\beta_1=\beta_2\neq\beta_3,\\
\beta_3=1,\\
 y_1+ y_2=0,\\
 y_3=-1.
\end{array}\right.
\end{equation}
The number of $(\beta_1,\beta_2,\beta_3,y_1,y_2,y_3)\in \Omega^3\times \left(\mathbb{F}_{3^m}^*\right)^3$ satisfying (\ref{eq system for N4 tran3}) is $3^m(3^m-1)$. If $y_1+y_2\neq0$, by arguments similar to Case 2, we can conclude that (\ref{eq system for N4 tran2}) has no solution.
Therefore, when $u_1=u_2\neq u_3$, (\ref{system matrix for r=4}) has
\begin{equation}\label{num solu for detB=0-case1}
3^m(3^m-1)
\end{equation}solutions.

Similarly, when $u_1=u_3\neq u_2$, or $u_2=u_3\neq u_1$, the number of solutions of (\ref{system matrix for r=4}) is also given by (\ref{num solu for detB=0-case1}).

\emph{Subcase (b):} $u_1=u_2=u_3$.  Then, (\ref{system matrix for r=4}) becomes
\begin{equation}\label{eq system for N4 tran5}\left \{
\begin{array}{lll}\beta_1=\beta_2=\beta_3,\\
\beta_1( y_1+ y_2+ y_3)=-1,\\
u_1^3\beta_1( y_1+ y_2+y_3)=-1,\\
\end{array}\right.
\end{equation}
which implies \begin{equation}\label{eq system for N4 tran6}\left \{
\begin{array}{lll}\beta_1=\beta_2=\beta_3=1,\\
 y_1+ y_2+ y_3=-1.\\
\end{array}\right.
\end{equation}
The number of $(\beta_1,\beta_2,\beta_3,y_1,y_2,y_3)\in \Omega^3\times \left(\mathbb{F}_{3^m}^*\right)^3$ satisfying (\ref{eq system for N4 tran6}) is $(3^m-1)+(3^m-2)^2.$

By Subcases (a) and (b), when  $\det(B)=0$, the number of solutions of (\ref{system matrix for r=4}) is given by \begin{equation}\label{num solu for detB=0}
3\cdot3^m(3^m-1)+(3^m-1)+(3^m-2)^2.
\end{equation}

\emph{Case \textrm{II}:} $\det(B)\neq 0$. Then, for each given $(\beta_1,\beta_2,\beta_3)\in \Omega^3$ such that $\det(B)\neq 0$, from (\ref{system matrix for r=4}), we can solve a unique solution $(y_1,y_2,y_3)$ as follows
\begin{equation*}\label{solution exp of y}\left \{
\begin{array}{lll}\beta_1y_1&=& -\frac{\left(u_2 - 1\right)\left(u_3 - 1\right)\left(u_2 + u_3 + 1\right)}{\left(u_1 - u_2\right)\left(u_1 - u_3\right) \left(u_1 + u_2 + u_3\right)},\\
\beta_2y_2&=& -\frac{\left(u_1 - 1\right)\left(u_3 - 1\right)\left(u_1 + u_3 + 1\right)}{\left(u_2 - u_1\right)\left(u_2 - u_3\right) \left(u_1 + u_2 + u_3\right)},\\
 \beta_3y_3&=& -\frac{\left(u_1 - 1\right)\left(u_2 - 1\right)\left(u_1 + u_2 + 1\right)}{\left(u_3 - u_1\right)\left(u_3 - u_2\right) \left(u_1 + u_2 + u_3\right)}.\\
\end{array}\right.
\end{equation*}
To ensure $(y_1,y_2,y_3)\in \left(\mathbb{F}_{3^m}^*\right)^3$,  we must have
\begin{equation}\label{trans-equation to cp}\left \{
\begin{array}{lll}u_1&=& \left(-\frac{\left(u_2 - 1\right)\left(u_3 - 1\right)\left(u_2 + u_3 + 1\right)}{\left(u_1 - u_2\right)\left(u_1 - u_3\right) \left(u_1 + u_2 + u_3\right)}\right)^{3^m-1},\\
u_2&=& \left(-\frac{\left(u_1 - 1\right)\left(u_3 - 1\right)\left(u_1 + u_3 + 1\right)}{\left(u_2 - u_1\right)\left(u_2 - u_3\right) \left(u_1 + u_2 + u_3\right)}\right)^{3^m-1},\\
 u_3&=&\left( -\frac{\left(u_1 - 1\right)\left(u_2 - 1\right)\left(u_1 + u_2 + 1\right)}{\left(u_3 - u_1\right)\left(u_3 - u_2\right) \left(u_1 + u_2 + u_3\right)}\right)^{3^m-1},\\
\end{array}\right.
\end{equation}
which implies
\begin{equation*}\left \{
\begin{array}{lll}(1-u_1)\left(u_1+u_2+u_3+1\right)^{3^m+1}= 1-u_1,\\
(1-u_2)\left(u_1+u_2+u_3+1\right)^{3^m+1}= 1-u_2,\\
 (1-u_3)\left(u_1+u_2+u_3+1\right)^{3^m+1}=1-u_3.\\
\end{array}\right.
\end{equation*}
Thus, (\ref{trans-equation to cp}) is equivalent to
\begin{equation}\label{final condition}\left \{
\begin{array}{lll}u_i\neq 1,\,\,i=1,2,3,\,\,\\
 u_1\neq u_2,\,\,u_2\neq u_3,\,\,u_1\neq u_3,\\
 u_1+u_2+u_3\neq 0,\\
 u_1+u_2+1\neq 0,\\
 u_1+u_3+1\neq 0,\\
 u_2+u_3+1\neq 0,\\
 u_1+u_2+u_3+1\in U.\\
\end{array}\right.\end{equation}Note that there is a one-to-one correspondence between $\beta\in \Omega$ and $u=\beta^{3^m-1}\in U$. Therefore, the above analysis shows that when $\det(B)\neq 0$, the number of solutions of (\ref{system matrix for r=4}) is the number of $(u_1,u_2,u_3)\in U^3$ satisfying (\ref{final condition}), which is exactly the number of $(u_1,u_2,u_3,u_4)\in U^4$ such that

\begin{equation}\label{final condition-tran}\left \{
\begin{array}{lll}u_i,\,\,i=1,2,3,4,\,\,{\rm and}\,\,1\,\,{\rm are}\,\,{\rm pairwise}\,\,{\rm distinct},\\
 u_1+u_2+u_3+u_4+1=0.\\
\end{array}\right.\end{equation}By Lemma \ref{five-sum-num}, the number of $(u_1,u_2,u_3,u_4)\in U^4$ satisfying (\ref{final condition-tran}) is \begin{equation}\label{num of solu for detBnot 0}\frac{5!B_5}{2^5(3^m+1)}.\end{equation}
By (\ref{num solu for detB=0}), (\ref{num of solu for detBnot 0}) and Lemma \ref{weightofzetterberg}, we have
\begin{equation}\label{value of N4'}N_4'=5\cdot3^n - 12\cdot3^m - (-1)^m 3^m + 9.\end{equation}
From (\ref{relation btw N4 and N4' }) and (\ref{value of N4'}), it follows the value of $N_4$.
\hfill$\square$
\begin{remark} The proof of Proposition  \ref{Npro} is similar to that of Proposition 2 in \cite{xiaieee2016},
where similar problem was considered over the finite field of characteristic $2$. It is interesting that for $p=2$ or $3$, the problem of determining $N_4$ over $\mathbb{F}_{p^n}$  can be reduced to the same combinatorial problem over $\mathbb{F}_{p^n}$ as stated in (\ref{final condition-tran}).
\end{remark}

\begin{remark} Compared with \cite{xiaieee2016}, the procedure for establishing the relation between the combinatorial problem (\ref{final condition-tran}) and the Zetterberg codes in this paper is more complicated. For $p=3$, the relation is not so obvious as that for $p=2$ and thus we have to carry out some analysis before obtaining relation (see Lemma \ref{five-sum-num} ). In addition, there are no explicit weight formulas available for the ternary Zetterberg codes and we need to establish some of these formulas. \end{remark}

With the above preparations, we can give the proofs of the main results now.

{\em Proof of Theorem \ref{exp-distribution theorem}:} By Lemma \ref{powersum} and Proposition \ref{Npro}, the power sum $\sum\limits_{(a,b)\in \mathbb{F}_{3^n}^2 }S(a,b)^r$ can be computed for $r=1,2,3,4$. Then,
\begin{equation*}\label{system of eq0}\left \{
\begin{array}{lll}\sum\limits_{i=0}^{4}\mu_i=3^{n}-1,\\
\sum\limits_{i=0}^{3}\left((i-1)3^m-1\right)^r\mu_i+\left(4\cdot3^m-1\right)^r\mu_4=\sum\limits_{(a,b)\in \mathbb{F}_{3^n}^2 }S(a,b)^r-\left(3^n-1\right)^r,\,\,r=1,2,3,4\end{array}\right.\end{equation*} forms  a system of five linear equations in five variables $\mu_i$, $i=0,1,2,3,4$. The coefficient matrix of this system is a Vandermonde matrix. By solving this equation system, $\mu_i$, $i=0,1,2,3,4$, are obtained. Thus, the value distribution of $S(a,b)$  defined (\ref{def of Sab}) is determined.

 \hfill$\square$

{\em Proof of Theorem \ref{cor-distribution theorem}:}  When $m\not\equiv2\,\,({\rm mod}\,\,4)$, $\gcd(d,3^n-1)=1$. Thus, $S(0,b)=-1$ for any $b\in \mathbb{F}_{3^n}^*$. We also have $S(a,0)=-1$ for any $a\in \mathbb{F}_{3^n}^*$, and $S(0,0)=3^n-1$. Then, by Theorem \ref{exp-distribution theorem}, the value distribution of $S(a,b)$ as $(a,b)$ runs through $\mathbb{F}_{3^n}^*\times \mathbb{F}_{3^n}^*$ can be calculated. Let $d^{-1}$ denote the inverse of $d$ modulo $3^n-1$. We have $S(a,b)=S(a/(-b)^{d^{-1}},-1)$ for $b\in \mathbb{F}_{3^n}^*$. Moreover, for each fixed $b\in \mathbb{F}_{3^n}^*$, the value distribution of $S(a/(-b)^{d^{-1}},-1)$ as $a$ runs through $\mathbb{F}_{3^n}^*$ is the same as  that of $C_d(\tau)$ when $\tau$ runs from $0$ to $3^n-2$.   Thus, the multiset $\{S(a,b) \mid (a,b) \in \mathbb{F}_{3^n}^*\times \mathbb{F}_{3^n}^*\}$ is the multiset sum of $3^n-1$ identical multisets $\{C_d(\tau) \mid \tau=0,1,\cdots,3^n-2\}$. By direct calculations, we get the desired result.
 \hfill$\square$

\section{Conclusion}\label{conclusion}
In this paper, for the ternary Niho decimation $d$ given in (\ref{def d}), the distribution of the cross-correlation function $C_d(\tau)$ defined in (\ref{dfC_d}) is derived. It is the first time that the correlation distribution for a non-binary Niho decimation has been determined since 1976. The main idea used here is similar to that in \cite{xiaieee2016} but the present paper involves more detailed computation techniques (see Remarks 1 and 2 at the end of Proposition \ref{Npro}).  When $p>3$, some conclusions presented in this paper no longer hold, and there is no result about the weight distribution of the $p$-ary Zetterberg code. Thus, when $p>3$, maybe new techniques are required to derive the correlation distribution for the Niho decimation (\ref{def d}).  The readers are invited to attack this problem.

\section*{Acknowledgment}
Y. Xia was supported by the National Natural Science Foundation of China under Grant
11301552. T. Helleseth and N. Li were
supported by the Norwegian Research Council. X. Zeng was supported
by the National Natural Science Foundation of China under Grant 61170257.

\section*{Appendix A}
Before we prove Lemma \ref{five-sum-num} in detail, we mention  the following properties of $\mathcal{T}_5$, which will be employed in the sequel.

\begin{lemma}\label{property of T5} Let $\mathcal{T}_5$ be defined by (\ref{k-tuple of U}) and $(t_1,t_2,t_3,t_4,t_5)$ be an element of $\mathcal{T}_5$. Define $t_k'=t_k+\varepsilon_k\frac{3^m+1}{2}\,\,({\rm mod}\,\,3^m+1)$, where $\varepsilon_k\in\{0,1\}$ and $k=1,2,3,4,5$. Then,

\noindent(\textrm{i}) the difference between any two of
$t_1$, $t_2$, $t_3$, $t_4$ and $t_5$ modulo $3^m+1$ cannot be $\frac{3^m+1}{2}$;

\noindent(\textrm{ii})
$\delta^{t_1'}+\delta^{t_2'}+\delta^{t_3'}+\delta^{t_4'}+\delta^{t_5'}=0$ if and only if $\varepsilon_1=\varepsilon_2=\varepsilon_3=\varepsilon_4=\varepsilon_5=0$ or $1$.
\end{lemma}
{\em Proof:}
(\textrm{i})  Suppose without loss of generality that $t_2\equiv t_1+\frac{3^m+1}{2}\,\,({\rm mod}\,\,3^m+1)$. Then, we have
$$\delta^{t_3}+\delta^{t_4}+\delta^{t_5}=0,$$
which leads to a contradiction due to Lemma \ref{threesum}. Thus, the desired result follows.

(\textrm{ii})
Obviously, the sufficient condition is true. It remains to prove the necessary condition. Suppose that $\varepsilon_k$, $k=1,2,3,4,5$, are not all equal. Then, assume without loss of generality that  $\varepsilon_1=\varepsilon_2=1$ and $\varepsilon_3=\varepsilon_4=\varepsilon_5=0$, which implies that
$$-\delta^{t_1}-\delta^{t_2}+\delta^{t_3}+\delta^{t_4}+\delta^{t_5}=0.$$ The above equation together with $\delta^{t_1}+\delta^{t_2}+\delta^{t_3}+\delta^{t_4}+\delta^{t_5}=0$ gives  $$\delta^{t_3}+\delta^{t_4}+\delta^{t_5}=0,$$ a contradiction to Lemma  \ref{threesum}. In the same way, other cases will also lead to a contradiction. Thus, the necessary  condition is also true.
\hfill$\square$

{\em Proof of Lemma \ref{five-sum-num}:} The conclusion of this lemma follows from the following three claims:

(\textrm{i}) from each element $(t_1,t_2,t_3,t_4,t_5)$ of $\mathcal{T}_5$ we can construct $2^5$ pairwise distinct codewords with weight five of $Z(q)$;

(\textrm{ii}) in (\textrm{i}) the codewords with weight five constructed from different elements of $\mathcal{T}_5$ are also pairwise distinct;

 (\textrm{iii}) each codeword with weight five in $Z(q)$ can be constructed from an element of $\mathcal{T}_5$.

\emph{Proof of Claim (\textrm{i}).} Let $(t_1,t_2,t_3,t_4,t_5)$ be an element in $\mathcal{T}_5$ and $\left(v_1,v_2,v_3, v_4,v_5\right)$  an arbitrary vector in $\left(\mathbb{F}_{3}^*\right)^5$.
Define $\varepsilon_k=1$ if $v_k=-1$ and $\varepsilon_k=0$ otherwise, and let
\begin{equation}\label{codeconstruct weight5}t_k'\equiv t_k+\varepsilon_k\frac{3^m+1}{2}\,\,({\rm mod}\,\,3^m+1),\end{equation}
where $1\leq k\leq 5$. Then, we have
\begin{equation}\label{converse-five-eq}v_1\delta^{t_1'}+v_2\delta^{t_2'}+v_3\delta^{t_3'}+v_4\delta^{t_4'}+v_5\delta^{t_5'}=0\end{equation}
since $\delta^{t_1}+\delta^{t_2}+\delta^{t_3}+\delta^{t_4}+\delta^{t_5}=0$ and $\delta^{\frac{3^m+1}{2}}=-1$. By Lemma \ref{property of T5} (\textrm{i}), $t_k'$, $1\leq k\leq 5$, in (\ref{converse-five-eq}) are also pairwise distinct. Let $\mathbf{c}=(c_0,c_1,\cdots,c_{q})$ be a vector defined by \begin{equation}\label{codeconstruct5}c_{t_k'}=v_k, \,\,k=1,2,\cdots,5\,\,{\rm and}\,\,c_j=0\,\,{\rm for}\,\,j\notin \{t_1',t_2',t_3',t_4',t_5'\}.\end{equation} Then, from (\ref{converse-five-eq}) and the definition of $Z(q)$ in (\ref{terZetterberg code}), one knows that the above vector $\mathbf{c}$ is a codeword of weight five in $Z(q)$. Thus, for a given element $(t_1,t_2,t_3,t_4,t_5)$ in $\mathcal{T}_5$, from each vector $\left(v_1,v_2,v_3, v_4,v_5\right)\in \left(\mathbb{F}_{3}^*\right)^5$ we can construct a codeword $\mathbf{c}$ having weight five in $Z(q)$ by (\ref{codeconstruct weight5}) and (\ref{codeconstruct5}).

Moreover, for a fixed element $(t_1,t_2,t_3,t_4,t_5)$ in $\mathcal{T}_5$, the codewords with weight five of $Z(q)$ constructed from different vectors of $\left(\mathbb{F}_{3}^*\right)^5$ are distinct. The reason is given as follows. Assume that $\left(v_{1,1},v_{1,2},v_{1,3}, v_{1,4},v_{1,5}\right)$ and $\left(v_{2,1},v_{2,2},v_{2,3}, v_{2,4},v_{2,5}\right)$ are two different vectors in $\left(\mathbb{F}_{3}^*\right)^5$.
Let $\varepsilon_{i,k}=1$ if $v_{i,k}=-1$ and $\varepsilon_{i,k}=0$ otherwise, and define  $t_{i,k}=t_k+\varepsilon_{i,k}\frac{3^m+1}{2}\,\,({\rm mod}\,\,3^m+1)$, where  $i=1,2$ and $k=1,2,3,4,5$.  Using the way described in (\ref{codeconstruct weight5}) and (\ref{codeconstruct5}), let the codeword constructed from $\left(v_{i,1},v_{i,2},v_{i,3}, v_{i,4},v_{i,5}\right)$ be $\mathbf{c}^i$, $i=1,2$. Suppose, on the contrary, that $\mathbf{c}^1=\mathbf{c}^2$. Then, we have
\begin{equation}\label{set in claim1}\{t_{1,k}\mid k=1,2,3,4,5\}=\{t_{2,k}\mid k=1,2,3,4,5\}.\end{equation}
From (\ref{set in claim1}), we can deduce that $t_{1,k}=t_{2,k}$ for each $k\in\{1,2,3,4,5\}$. Otherwise assume that $t_{1,i}=t_{2,j}$ for some  $i,\,j\in\{1,2,3,4,5\}$ and $i\neq j$. Then, we conclude that $\left(t_i-t_j\right)\,\,({\rm mod}\,\,3^m+1)$ is equal to $0$ or $\frac{3^m+1}{2}$, a contradiction to Lemma \ref{property of T5} (\textrm{i}). Since $t_{1,k}=t_{2,k}$ for each $k\in\{1,2,3,4,5\}$,  we must have  $\varepsilon_{1,k}=\varepsilon_{2,k}$, i.e., $v_{1,k}=v_{2,k}$ for each $k\in\{1,2,3,4,5\}$, a contradiction to the assumption that $\left(v_{1,1},v_{1,2},v_{1,3}, v_{1,4},v_{1,5}\right)$ and $\left(v_{2,1},v_{2,2},v_{2,3}, v_{2,4},v_{2,5}\right)$ are different.

Moreover, note that the total number of vectors in $\left(\mathbb{F}_{3}^*\right)^5$ is $2^5$. From the above discussion, it follows Claim (\textrm{i}).

  \emph{Proof of Claim (\textrm{ii}).}   Let $(\tau_1,\tau_2,\tau_3,\tau_4,\tau_5)$ be another element of $\mathcal{T}_5$, which is different from $(t_1,t_2,t_3,t_4,t_5)$.  Let $\left(v_1,v_2,v_3, v_4,v_5\right)$ and $\left(w_1,w_2,w_3, w_4,w_5\right)$ are two vectors in $\left(\mathbb{F}_{3}^*\right)^5$. Define $\eta_k=1$ if $w_k=-1$ and $\eta_k=0$ otherwise, and let
$\tau_k'\equiv \tau_k+\varepsilon_k\frac{3^m+1}{2}\,\,({\rm mod}\,\,3^m+1)$, $1\leq k\leq 5$. Using the method in (\ref{codeconstruct weight5}) and (\ref{codeconstruct5}) and the notation there, let $\mathbf{c}$ be the codeword  constructed from $(t_1,t_2,t_3,t_4,t_5)$ and $\left(v_1,v_2,v_3, v_4,v_5\right)$, and $\mathbf{c}'=(c_0',c_1',\cdots,c_{3^m}')$ the codeword constructed from $(\tau_1,\tau_2,\tau_3,\tau_4,\tau_5)$ and $\left(w_1,w_2,w_3, w_4,w_5\right)$.  We assume $\mathbf{c}'=\mathbf{c}$ and obtain a contradiction.

From $\mathbf{c}'=\mathbf{c}$, we have \begin{equation*}\label{setequalclaim2}\{t_k'\mid k=1,2,3,4,5\}=\{\tau_k'\mid k=1,2,3,4,5\}\end{equation*}
which implies that  for each $i\in \{1,2,3,4,5\}$, there exists a unique $\pi(i)\in \{1,2,3,4,5\}$ such that \begin{equation}\label{relation of coefclaim2}\tau_i+\eta_i\frac{3^m+1}{2}\equiv t_{\pi(i)}+\varepsilon_{\pi(i)}\frac{3^m+1}{2}\,\,({\rm mod}\,\,3^m+1),\end{equation} where $\pi$ is a permutation of the set $\{1,2,3,4,5\}$. We rewrite (\ref{relation of coefclaim2}) as

\begin{equation}\label{relation of coefclaim2-1}\tau_i\equiv t_{\pi(i)}+\left(\varepsilon_{\pi(i)}-\eta_i\right)\frac{3^m+1}{2}\,\,({\rm mod}\,\,3^m+1)\end{equation}
for each $i\in\{1,2,3,4,5\}$.
Note that both $(t_1,t_2,t_3,t_4,t_5)$ and $(\tau_1,\tau_2,\tau_3,\tau_4,\tau_5)$ are in $\mathcal{T}_5$. By (\ref{relation of coefclaim2-1}) and Lemma \ref{property of T5} (\textrm{ii}), we have \begin{equation}\label{final for claim 2}\varepsilon_{\pi(i)}-\eta_i\equiv 1\,\,({\rm mod}\,\,2)\,\,{\rm for}\,\,{\rm each}\,\,i\in\{1,2,3,4,5\}.\end{equation}
Furthermore, if $\mathbf{c}'=\mathbf{c}$, by  (\ref{relation of coefclaim2}) we also have
$c'_{\tau_i'}=w_i=c_{t_{\pi(i)}'}=v_{\pi(i)}$ for each $i\in\{1,2,3,4,5\}$,  which contradicts with  (\ref{final for claim 2}).  Thus, $\mathbf{c}'\neq \mathbf{c}$ and Claim (\textrm{ii}) holds.

\emph{Proof of Claim (\textrm{iii}).}
Let $\mathbf{c}=(c_0,c_1,\cdots,c_{q})$ be a codeword of weight five in $Z(q)$ and assume that $c_{i_1},c_{i_2},c_{i_3}, c_{i_4},c_{i_5}\in \mathbb{F}_{3}^*$, $0\leq i_1<i_2<i_3<i_4<i_5<3^m+1$ and $c_j=0$ for $j\notin \{i_1,i_2,i_3,i_4,i_5\}$.  According to the definition of $Z(q)$, we must have
\begin{equation}\label{initial-c-equal}c_{i_1}\delta^{i_1}+c_{i_2}\delta^{i_2}+c_{i_3}\delta^{i_3}+c_{i_4}\delta^{i_4}+c_{i_5}\delta^{i_5}=0.\end{equation}
 Let $i_k'=i_k+\frac{3^m+1}{2}\,\,({\rm mod}\,\,3^m+1)$ if $c_{i_k}=-1$ and $i_k'=i_k$ otherwise, $k=1,2,3,4,5$. Then, (\ref{initial-c-equal}) can be rewritten as
\begin{equation}\label{c-equal1}\delta^{i_1'}+\delta^{i_2'}+\delta^{i_3'}+\delta^{i_4'}+\delta^{i_5'}=0.\end{equation}  In the sequel we will show that $i_k'$ in (\ref{c-equal1}), $1\leq k\leq 5$, are pairwise distinct, and then by sorting them in increasing order,  we get an element of  $\mathcal{T}_5$.  From this element of $\mathcal{T}_5$ and $(c_{i_1},c_{i_2},c_{i_3},c_{i_4},c_{i_5})$, one can see that the codeword $\mathbf{c}$ given here can be constructed via the method described in (\ref{codeconstruct weight5}) and (\ref{codeconstruct5}).

 Now we turn to proving that $i_k'$, $1\leq k\leq 5$, in (\ref{c-equal1}) are pairwise distinct. We assume without loss of generality that  at most two of $c_{i_1},c_{i_2},c_{i_3}, c_{i_4}$ and $c_{i_5}$ in (\ref{initial-c-equal}) take the value $-1$. Otherwise, replace $(c_{i_1},c_{i_2},c_{i_3}, c_{i_4}, c_{i_5})$ by $(-c_{i_1},-c_{i_2},-c_{i_3}, -c_{i_4}, -c_{i_5})$ in (\ref{initial-c-equal}). Thus, we only need to consider the following two cases.

\textit{Case 1:} Only one of $c_{i_1}, c_{i_2}, c_{i_3}, c_{i_4}$ and $c_{i_5}$ takes the value $-1$.
Without loss of generality, we assume that $c_{i_1}=-1$. Then, (\ref{c-equal1}) becomes
\begin{equation}\label{c-equal2}\delta^{i_1'}+ \delta^{i_2}+\delta^{i_3}+\delta^{i_4}+\delta^{i_5}=0,\end{equation}
where $i_1'=i_1+\frac{3^m+1}{2}\,\,({\rm mod}\,\,3^m+1)$, and we only need to show that none of $i_2$, $i_3$, $i_4$ and $i_5$ are equal to $i_1'$. Without loss of generality, suppose that $i_2=i_1'$. Then, $\delta^{i_1'}+ \delta^{i_2}=\delta^{i_1}$ and (\ref{c-equal2}) can be rewritten as
 \begin{equation}\label{c-equal3}\delta^{i_1}+ \delta^{i_3}+\delta^{i_4}+\delta^{i_5}=0.\end{equation}
 According to Lemma \ref{four U}, (\ref{c-equal3}) holds if and only if $$0\leq i_1<i_3<\frac{3^m+1}{2}, \,\,i_4=\frac{3^m+1}{2}+i_1\,\,{\rm and}\,\,i_5=\frac{3^m+1}{2}+i_3.$$
Thus, we have $i_2=i_4$, a contradiction to the assumption $0\leq i_1<i_2<i_3<i_4<i_5<3^m+1$. Therefore, in this case we have proved that $i_k'$, $1\leq k\leq 5$, in (\ref{c-equal1}) are pairwise distinct.

\textit{Case 2:} Two of $c_{i_1}, c_{i_2}, c_{i_3}, c_{i_4}$ and $c_{i_5}$ take the value $-1$. Without loss of generality, we assume that
$c_{i_1}=c_{i_2}=-1$.
  Then, (\ref{c-equal1}) can be rewritten as
  \begin{equation}\label{c-equal4}\delta^{i_1'}+ \delta^{i_2'}+\delta^{i_3}+\delta^{i_4}+\delta^{i_5}=0,\end{equation}
where $i_k'=i_k+\frac{3^m+1}{2}\,\,({\rm mod}\,\,3^m+1)$, $k=1,2$. Note that $i_1'\neq i_2'$ and $i_3<i_4<i_5$. Thus, the number of distinct values in the multi-set $\{i_1', i_2', i_3, i_4, i_5\}$ is at least $3$. To prove $i_1', i_2', i_3, i_4$ and $i_5$ in (\ref{c-equal4}) are pairwise distinct, we need to exclude the following two cases.

\textrm{(a)} The number of distinct values in the multi-set $\{i_1', i_2',i_3,i_4,i_5\}$ is $4$. Without loss of generality, suppose that $i_1'=i_3$. Then, (\ref{c-equal4}) becomes
\begin{equation}\label{c-equal5}\delta^{i_1}+ \delta^{i_2'}+\delta^{i_4}+\delta^{i_5}=0.\end{equation} Then, by a discussion similar to Case 1, (\ref{c-equal5}) also leads to a contradiction. Thus, case \textrm{(a)} cannot occur.

\textrm{(b)} The number of distinct values in the set $\{i_1', i_2',i_3,i_4,i_5\}$ is $3$. Without loss of generality, we assume $i_1'=i_3$ and $i_2'=i_4$. Then, (\ref{c-equal4}) becomes
\begin{equation*}\delta^{i_1}+ \delta^{i_2}+\delta^{i_5}=0,\end{equation*}
which cannot hold due to Lemma \ref{threesum}, and thus case \textrm{(b)} cannot occur either.

Combining the discussion in Case $1$ and $2$, we conclude that $i_k'$, $1\leq k\leq 5$, in (\ref{c-equal1}) are pairwise distinct.
\hfill$\square$

\end{document}